\documentclass[12pt,preprint]{aastex}

\shorttitle{Infrared observations of a mid-L dwarf with strong H$\alpha$ emission}
\shortauthors{Riaz \& Gizis}

\begin{document}

\title{Infrared observations of a mid-L dwarf with strong H$\alpha$ emission}

\author{Basmah Riaz and John E. Gizis}
\affil{Department of Physics and Astronomy, University of Delaware,
    Newark, DE 19716; basmah@udel.edu, gizis@udel.edu}

\begin{abstract}
We present {\it Spitzer}/IRAC observations of the L5 dwarf, 2MASSI J1315309-264951 (2M1315). This ultracool dwarf is known to display strong emission in the H$\alpha$ line. The SED for this object does not show any IR excess, that would indicate the presence of an accretion disk. Although the IRAC colors for 2M1315 are consistent with other L dwarfs, they seem to be redder by $\sim$0.1 mag compared to the other L5 dwarfs, and more like the late-type L dwarfs. The existing six epochs of spectroscopy suggest that the emission in H$\alpha$ is not persistent, but shows long-term variability between a flare value of $\sim$100 $\AA$ and a quiescent value of $\sim$25 $\AA$. Chromospheric activity seems to be the most likely cause, which is also indicated by the detection of  Na I D lines in emission (Fuhrmeister et al.). We have measured a proper motion of 0.79$\arcsec\pm$ 0.06$\arcsec$/yr, that corresponds to a tangential velocity of $\sim$81 km/s, at a distance of $\sim$22 pc. The high $V_{tan}$ for this object suggests an old age. Evolutionary models indicate lower limits of 3.3 Gyr (Chabrier et al.) or 1.4 Gyr (Burrows et al.) for this magnetically active L dwarf to be a stable hydrogen-burning star. There seem to be no observational differences between this old L dwarf that has H$\alpha$ emission and the other field L dwarfs without it. 
\end{abstract}


\section{Introduction}

Emission in H$\alpha$ is commonly seen among M dwarfs, with the activity levels increasing towards later types due to longer spin-down timescales, reaching a peak at M7. Beyond spectral type M7, there is a drop off in H$\alpha$ activity, with a rapid decline seen for L dwarfs (Gizis et al. 2000). This suggests a decrease in the generation and dissipation of magnetic fields for late-M and L dwarfs. One reason could be the increasing atmospheric neutrality in ultracool dwarfs that decouples the field from atmospheric motions, and can hamper their dissipation, even if strong fields are generated (Mohanty et al. 2002). This decoupling can cause suppression of chromospheric and coronal heating, resulting in a decrease in emission in the H$\alpha$ line and X-rays, despite the presence of magnetic fields. Indeed, magnetic field strengths of 0.1-1kG have been measured for some late-M and L dwarfs (e.g. Berger 2006) indicating that the drop in H$\alpha$ emission seen for ultracool dwarfs must be due to atmospheric neutrality rather than a fall in magnetic flux production (Reiners \& Basri 2006). 

On the contrary, there exists a small population of late-M and L dwarfs, such as PC0025+0447 (M9.5; Schneider et al. 1991), 2MASSW J0149090+295613 (M9.5; Liebert et al. 1999) and 2MASS J01443536-0716142 (L5; Liebert et al. 2003), that exhibit strong flaring in the H$\alpha$ emission line. This flaring is occasional and occurs only on the order of $\sim$1\% of the time (Liebert et al. 2003). While different explanations have been suggested for these rare displays of strong H$\alpha$ emission, the mechanism that empowers such flares is still unknown. We report here on {\it Spitzer} (Werner et al. 2004) IRAC (Fazio et al. 2004) observations of the mid-type L dwarf 2MASSI J1315309-264951 (hereafter, 2M1315). 2M1315 was found to show strong emission in H$\alpha$ by Hall (2002a), that decreased by a factor of 2 half a year later (from an equivalent width (EW) of 121$\pm$31$\AA$ to 25$\pm$10$\AA$). Hall classified this object as an L3 dwarf, and concluded that the variable H$\alpha$ emission is either due to chromospheric activity, or by accretion in a binary system. Gizis (2002) independently found 2M1315 and classified it as an L5 dwarf. However, he found this object to be a persistent strong H$\alpha$ emitter (EW =97$\pm10\AA$) from 6 consecutive 600s exposures on one night, and 2 consecutive 900s exposures the following night. On the basis of an absence of low surface gravity signatures in its spectrum, Gizis concluded that this object is not a member of TWA, and may be a relatively old L dwarf. This was confirmed by Hall (2002b) who found 2M1315 to have a high proper motion (0.71$\arcsec \pm$ 0.07$\arcsec$/yr), that corresponds to a $V_{tan}$ of $\sim$76 km/s at d$\sim$23 pc. Hall (2002b) again found a large EW of 124 $\AA$, and thus concluded that 2M1315 is a persistent strong H$\alpha$ emitter. Later, Fuhrmeister et al. (2005) reported the weak detection of Na I D lines in emission, and an H$\alpha$ EW of 24.1 $\AA$, suggesting that 2M1315 was in a quiescent state since Hall's measurement. Recently, Barrado y Navascu\'{e}s (2006) has reported an average EW of 153$\pm$26 $\AA$ from seven consecutive spectra obtained in March 2003.

The likely explanations for strong H$\alpha$ emission can be narrowed down to (i) presence of an unusually strong magnetic field, (ii) accretion in a closely interacting brown dwarf binary system, (iii) presence of an accretion disk. We have obtained IRAC observations of 2M1315 in order to check for IR excess, that would indicate the presence of a disk around it. In Section~\ref{obs}, we present our observations. Section \ref{disc} gives a discussion of the previous six epochs of spectroscopy, and the likely causes of strong H$\alpha$ emission seen in this L dwarf.

\section{Observations}
\label{obs}

Observations in the IRAC bands were obtained in February, 2006, using the full array mode. A frame time of 12 sec was requested, along with a dither pattern of five positions Gaussian, which allows the removal of instrumental artifacts. Aperture photometry was performed using the task PHOT under the IRAF APPHOT package. An aperture radius of 5 pixels and a background annulus of 10-20 pixels was used. The zero point fluxes of 280.9, 179.7, 115 and 64.1 Jy and aperture corrections of 1.05, 1.05, 1.058 and 1.068 were used for IRAC channels 1 through 4, respectively. The errors in magnitudes are between $\sim$0.01 and 0.02 mag, with an additional IRAC calibration uncertainty of $\sim$2\%. Table 1 lists the observed fluxes in the IRAC bands.

\section{Discussion}
\label{disc}

In order to determine the extent of IR excess, we have looked at different models for very low-mass stars so as to obtain a good fit to the photospheric emission from 2M1315. The models by Allard et al. (2001) are two limiting cases of dust treatment in the atmosphere. The DUSTY models include not only grain formation, but also grain opacities. As first shown by Tsuji et al. (1996), dust opacities significantly affect the emergent flux from an ultracool dwarf, as they deplete the gas phase of a number of molecular species, such as, TiO, VO, FeH, CaH, MgH, and of refractory elements, such as Al, Ca, Ti and Fe. The resulting spectral distribution thus appears redder with weaker molecular opacities, as compared to the models that do not consider grain opacities. The COND models by Allard et al. (2001), do consider grain formation but do not take into account the grain opacities, resulting in a spectral distribution that is transparent blueward of 1.0$\micron$, and very similar to the dust-free case (the NextGen models; Hauschildt et al. 1999a,b). As discussed in Chabrier et al. (2000; hereafter C00), in the DUSTY models, turbulent mixing is very efficient in bringing material upward to the region of condensation and maintaining small grain layers, which would otherwise gravitationally settle down below the photosphere. In the COND model, however, the turbulent mixing cannot balance gravitational settling, which results in the sedimentation of all of the condensed species below the photosphere. Intermediate between the two extremes of DUSTY and COND are the SETTLE models by Allard et al. (2003). These models assume a sedimentation efficiency higher than the DUSTY but lower than the COND models, i.e. when grains have partially settled below the photosphere.

A spectral type of L5 (Gizis 2002) suggests an effective temperature of $\sim$1700 K (Kirkpatrick 2005) for 2M1315. In Fig. 1a, we have used the DUSTY, SETTLE and COND models to get a good fit to the photospheric emission from 2M1315. The COND models result in bluer colors for decreasing $T_{eff}$, and more flux shortward of 1$\micron$, as compared to the DUSTY ones, while the SETTLE models show fluxes intermediate between the two. Fig. 1a shows the observed near-IR flux to be over-estimated by the COND and SETTLE models for 2M1315, whereas the DUSTY one provides a good fit. Allard et al. (2001) have discussed the affects of gravity on the DUSTY models (see their Figs. 16 and 17). Due to the higher pressures and densities of the high-gravity models, the optical to red spectral region is significantly veiled by dust opacities. The near-IR water vapor bands are also affected. A log g=5.0 DUSTY model underestimates the observed flux in the J-band for 2M1315, as can be seen from Fig. 1b, whereas log g=4.0 provides a good fit. We note that a lower value of log g does not imply that 2M1315 is a relatively young star. The temperature 1700 K lies at the cut-off between the DUSTY and the COND models, suggesting that the models may be inconsistent at this temperature. The observed (J-K) colors for the reddest L dwarfs show a saturation at a value of $\sim$1.9, corresponding to a $T_{eff}\sim$1700-1800 K (see Fig. 6 in C00). 2M1315 lies in this region, with a (J-K) color of 1.7 and an absolute K magnitude of 11.8 (Fig. 1c). Due to the continued influence of dust opacities, the DUSTY models continue to redden to colors (J-K)$>$2, and thus fail to reproduce the colors for mid- to late-type L dwarfs ($M_{K}\ga$11.5, J-K $\ga$1.5). C00 have explained this to be caused by the gravitational settling of some grains below the photosphere. Fig. 1c shows that while the SETTLE models can reproduce the turn to the blue of (J-K) colors that is seen from mid-L to T-types in color-magnitude diagrams, they still fall short of reproducing the colors of the reddest L dwarfs. The cloudy models by Marley et al. (2002) which, similar to the SETTLE case, assume a sedimentation efficiency intermediate between the DUSTY and COND models, have been able to reproduce the (J-K) colors for the latest type L dwarfs.

The SED for 2M1315 (Fig. 1a) does not show any significant excess emission in any of the IRAC bands. The 3.6 to 8 $\micron$ slope for 2M1315 is $\sim$-2.5. According to the criteria defined in Lada et al. (2006), this makes it a disk less object, consistent with our finding. The main spectral features in the IRAC bands are due to molecular absorption of $CH_{4}$ near 3.3 and 7.8 $\micron$, and $H_{2}O$ at 5.5 and 6.5$\micron$ (Noll et al. 2000, Roelling et al. 2004, Cushing et al. 2006). The absorption in these bands strengthens towards later types. The dip seen in 2M1315's SED at 4.5$\micron$ could be due to the CO band at 4.7$\micron$, as seen in the T dwarf GL 229B (Oppenheimer et al. 1998).

In Fig. 2, we have compared the IRAC colors of 2M1315 with other M, L and T dwarfs from Patten et al. (2006). Although the colors for 2M1315 are consistent with other L dwarfs, they seem to be redder by $\sim$0.1 mag compared to the other L5 dwarfs, and more like the late-type L dwarfs. More CO absorption near 4.5 $\micron$ relative to $H_{2}O$ near 5.5 $\micron$ can explain the redder [4.5]-[5.8] color. A redder [3.6]-[8] color suggests more absorption of methane near 3.3$\micron$, compared to 7.8$\micron$. Could a (slightly) redder color suggest a higher surface gravity for 2M1315? To answer this, we have checked for any variations in {\it Spitzer} colors with surface gravity or tangential velocity. In Fig. 2, L dwarfs from Patten et al. for which Li EW has been measured are denoted by filled diamonds. For the others, Li EW could not be found in the literature. We have also determined the tangential velocities for M and L dwarfs in Patten's sample by searching for their proper motion and parallaxes in the SIMBAD database. We could only find proper motions for 24 objects; out of these, six have $V_{tan}$ between 50 and 80 km/s, and are denoted by filled circles. Fig. 2 shows that the IRAC colors for M and L dwarfs are not very sensitive to surface gravity, or tangential velocities. Though the number of data points used for comparison is very small, the lack of any concentration of the higher velocity (older) objects near 2M1315, or that of the objects with lithium (lower surface gravity, younger) near bluer IRAC colors suggests that these colors are not very useful as age indicators.

From the 2MASS and IRAC images, we have determined the proper motion for 2M1315 to be 0.79$\arcsec \pm$ 0.06$\arcsec$/yr ($\mu_{\alpha}$ = -0.73$\arcsec$/yr, $\mu_{\delta}$ = -0.31$\arcsec$/yr). Earlier, Hall (2002b) had found a proper motion of 0.71$\arcsec \pm$0.07$\arcsec$/yr, assuming that the nearby bright star, USNO J131531.230-264953.01, that lies at a separation of 7.14$\arcsec$ $\pm$ 0.05$\arcsec$ (epoch 2002.685) has negligible proper motion. Comparing the 2MASS and IRAC images shows that the position of the USNO star has not changed, and thus confirms that it is 2M1315 that exhibits the high proper motion. A distance of 21.7 pc (obtained from the relation in Dahn et al. 2002) results in a tangential velocity of 81.3 km/s. The proper motions for ten known TWA members range between 0.06 and 0.13$\arcsec$/yr, resulting in $V_{tan}$ between 13 and 32 km/s (using distances from Reid 2003). This rules out a TWA membership for 2M1315, and suggests an old age. That 2M1315 is relatively old and massive was suggested earlier by Gizis (2002) and Hall (2002b), based on the absence of low surface gravity signatures and its high proper motion, respectively. Fig. 3 shows the age isochrones for masses 0.09$M_{\sun}$ (top) to 0.02$M_{\sun}$ (bottom), from C00 DUSTY (bold lines), COND (long-dashed lines), and Burrows et al. (1997; hereafter B97) (short-dashed lines) models. The B97 models are updated versions of their 1995 models, and show a change of less than 10\% in luminosity at a given age and mass. C00 have discussed that the number of grain species in the B97 model is very small, and the grain opacities are treated as frequency-independent, resulting in a gray atmosphere structure above 1300K. C00 models consist of a significant number of grain species, the opacities of which are frequency-dependent, resulting in nongray atmosphere for 900 $\la$ $T_{eff}$ $\la$ 2800K. Fig. 3 suggests that the decrease in effective temperatures due to silicate formation between $10^8$ and $10^9$ yrs is steeper in B97 tracks, as compared to C00 models. This could explain the lower effective temperatures seen for B97 tracks during the H-burning phase. At a given age and mass, the temperature difference between the two models can be as large as $\sim$400K, and seems to be more pronounced for the higher masses. The H-burning limit for the DUSTY models is 0.07$M_{\sun}$. C00 had found that including dust opacities lowers the effective temperature and luminosity for a given mass. The difference in $T_{eff}$ between the DUSTY and COND models is more obvious at lower temperatures, and could reach up to $\sim$10\%. The COND models thus have a higher H-burning limit at 0.072$M_{\sun}$. The Li-burning limit for a 1 Gyr object from DUSTY models is 0.055$M_{\sun}$. For higher masses, the Li abundance goes to zero. Hall (2002b) had suggested an age of $\ga$ 2 Gyr for 2M1315. For an effective temperature of 1700 K, 2M1315 would have to be younger than $\sim$760 Myr (C00) or $\sim$630 Myr (B97) to be a Li-burning brown dwarf. For it to be a H-burning star, the lower age limits are 3.3 Gyr (C00) or 1.4 Gyr (B97). The strong H$\alpha$ emission shown by this likely H-burning star is in agreement with the results from Gizis (2000), who concluded that some active L dwarfs are drawn from an older, more massive population. He has also noted the presence of another population of L dwarfs like Kelu 1, that show both H$\alpha$ emission and Li absorption. Given 2M1315's old age, its emission can be related to having a higher surface gravity and a fusion energy source.

There are now six reported measurements of H$\alpha$ EW for 2M1315. These are listed in Table 2, along with the dates of observation and the spectral resolution. Hall (2002a) have discussed that their 2001 August EW measurement of 25 $\pm$ 10 $\AA$ is an underestimate as that spectrum may have been contaminated by the flux from the nearby USNO star. Thus discarding this measurement, he concluded that 2M1315 shows persistent non-variable H$\alpha$ emission of EW $\sim$ 100 $\AA$, with log ($L_{H\alpha}$/$L_{bol}$) $\sim$ -4. However, the EW measurement of 24.1 $\AA$ by Fuhrmeister et al. (2005) is from a $\sim$45,000 resolution spectrum, and indicates that the H$\alpha$ emission seen for 2M1315 may not be persistent, but variable. The values in Table 2 suggest a long-term variation in H$\alpha$ emission, with an EW of $\sim$25 and $\sim$100 $\AA$ defining the quiescent and the maximum states, respectively. We note that the H$\alpha$ EW depends strongly on the resolution, as discussed in White \& Basri (2003). At low resolution, the 6569 $\AA$ TiO feature blends with the H$\alpha$ line, resulting in an underestimate of the continuum redward of the H$\alpha$ feature, and an overestimate of the H$\alpha$ EW. The higher the resolution, the smaller the EW, with the changes being as large as $\sim$30\% in going from low to high resolution.

Our {\it Spitzer} observations can now rule out the presence of an accretion disk around 2M1315, which could have explained its strong H$\alpha$ emission. An absence of IR excess was, however, expected, considering the conclusions made by Gizis (2002) and Hall (2002b) about its age. An age of a Gyr or more makes 2M1315 unlikely to be a T Tauri analog brown dwarf, and is old enough to sweep away all signs of a debris disk around it.

The other possible explanation for H$\alpha$ emission in 2M1315 could be the accreting brown dwarf binary scenario, discussed further in Burgasser et al. (2000). However, as Hall (2002a) suggested, this would require some variability in the accretion from the sustained Roche lobe flow, that would explain the variability in H$\alpha$ emission.

The most likely explanation seems to be chromospheric activity. The long-term variations in H$\alpha$ suggest a slowly varying magnetic field, the strength of which though must be much stronger than that for other L and T dwarfs. This can be seen from Fig. 2 in Liebert et al. (2003), where even the quiescent emission of 2M1315 is higher by $\sim$2 orders of magnitude than the upper limits for other L dwarfs at the same spectral subclass. Hall (2002a) had suggested that the variable H$\alpha$ emission is more likely to be due to rapid variations, like flaring, than slow variations in the magnetic field strength, since this is rare in late-type dwarfs. However, flaring would require a shorter timescale for the variations to occur than that seen for 2M1315. Fuhrmeister et al. (2005) reported weak detection of Na I D lines in emission in 2M1315, the first such detection in an L dwarf. This again suggests that 2M1315 is a chromospherically active object. Comparing with the other two very late-type field dwarfs, PC 0025+0447 and 2MASSI J1237392+652615, that show persistent strong H$\alpha$ emission, Liebert et al. (2003) had concluded that these, along with 2M1315, could either be young (10-100 Myr) rare objects in the solar neighborhood, or a $\arcsec$special$\arcsec$, yet unknown dynamo could be responsible for generating a stronger-than-average magnetic field in these objects. Such a possibility is discussed in Mullan \& MacDonald (2001), who have considered several nonstandard additions to the $\arcsec$standard models$\arcsec$ of main-sequence stars, in order to explain why the magnetically active and inactive stars form two distinct populations in a $T_{eff}$-radius diagram. Their models suggest that the onset of a radiative core in an object of mass as low as 0.15$M_{\sun}$ requires $T_{eff}$ to drop to a value of $\sim$1500 K, and stronger magnetic fields with upper limits on strength of about 100 MG. In other words, a strong magnetic field could make a H-burning star look like a cool L dwarf. Mullan \& MacDonald have also suggested that at a given $T_{eff}$, active stars have larger radii than inactive ones. 2M1315 could be more luminous due to its stronger activity, if these nonstandard models are correct. 

We also consider the possibility that if the inclusion of dust opacities significantly influences the emergent flux, it might also affect the chromospheric line emission. In their model chromospheres for mid- to late-type M dwarfs, Fuhrmeister et al. (2005) have compared the H$\alpha$ line emission for the models with and without dust (see their Fig. 3). These authors have found that including dust opacities enhances most of the emission lines and lowers the continuum. The flux in the H$\alpha$ line in their models increased by a factor of $\sim$1.3 by including dust opacities. Hall (2002a) had found that the H$\alpha$ flux in their 2001 March measurement (EW = 121 $\pm$ 31 $\AA$) for 2M1315 was a factor of 2 higher than the 2001 August measurement (EW = 25 $\pm$ 10 $\AA$). Thus dust opacities could affect the chromospheric line emission, and consequently the EW measurement.

In summary, the H$\alpha$ emission in 2M1315 is not persistent, but displays a long-term variation between a flare value of $\sim$100 $\AA$ and a quiescent value of $\sim$25 $\AA$. The absence of circumstellar material supports the chromospheric model. The magnetic field heating up the chromosphere must be unusually strong, but slowly varying. It is though not clear why this object has strong H$\alpha$ emission from the chromosphere, while other old L dwarfs do not. 2M1315 has a high proper motion (0.79$\arcsec \pm$ 0.06$\arcsec$/yr) which results in a $V_{tan}$ of $\sim$81 km/s at a distance of $\sim$22 pc. The high velocity suggests that an old age is likely, and rules out a TWA membership. At an effective temperature of 1700 K, evolutionary models indicate lower limits of 3.3 Gyr (C00) or 1.5 Gyr (B97) for this magnetically active brown dwarf to be a hydrogen-burning star.

\acknowledgments
This work is based on observations made with the Spitzer Space Telescope, which is operated by the Jet Propulsion Laboratory, California Institute of Technology under a contract with NASA. Support for this work was provided by NASA through an award issued by JPL/Caltech. This work has made use of the SIMBAD database.

\clearpage
\begin{figure}
 \begin{center}
    \begin{tabular}{cc}
      \resizebox{80mm}{!}{\includegraphics[angle=0]{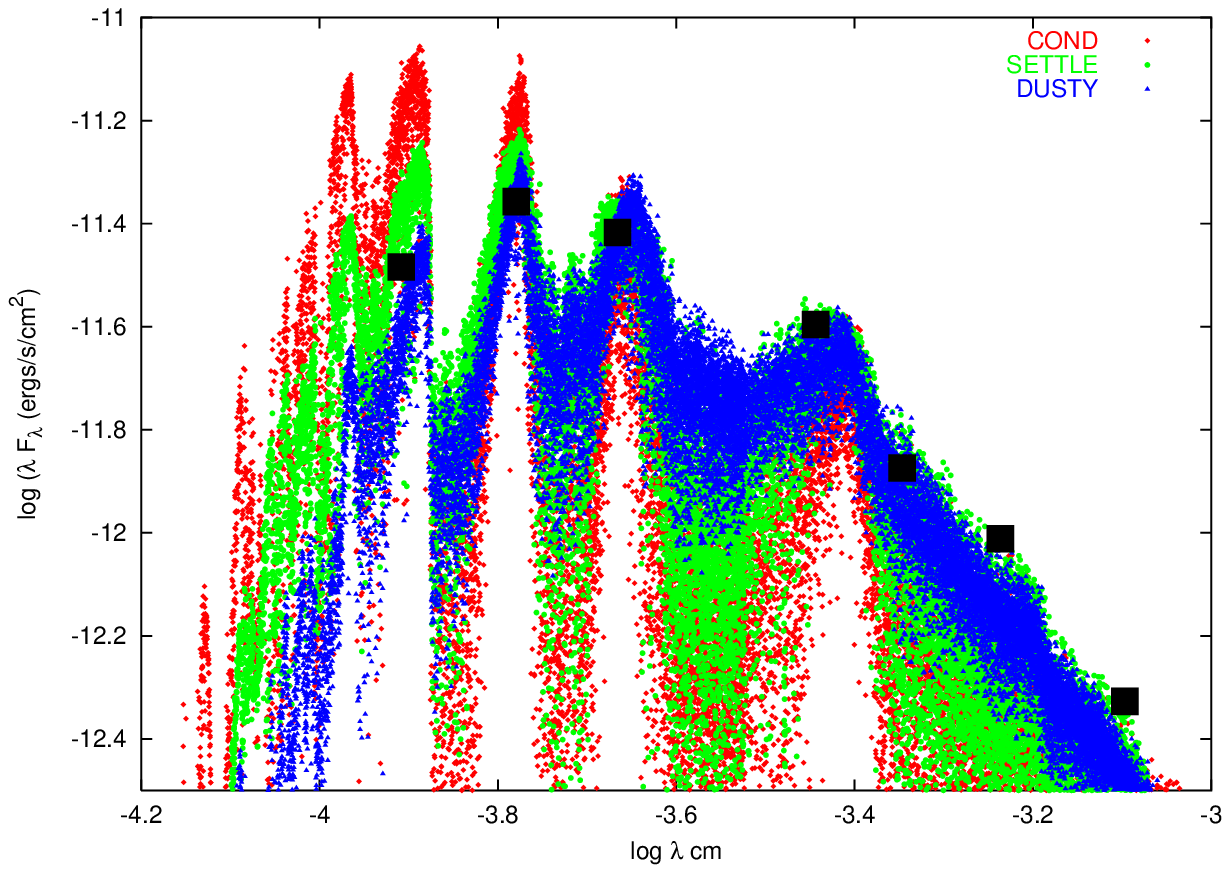}} \\
      \resizebox{80mm}{!}{\includegraphics[angle=0]{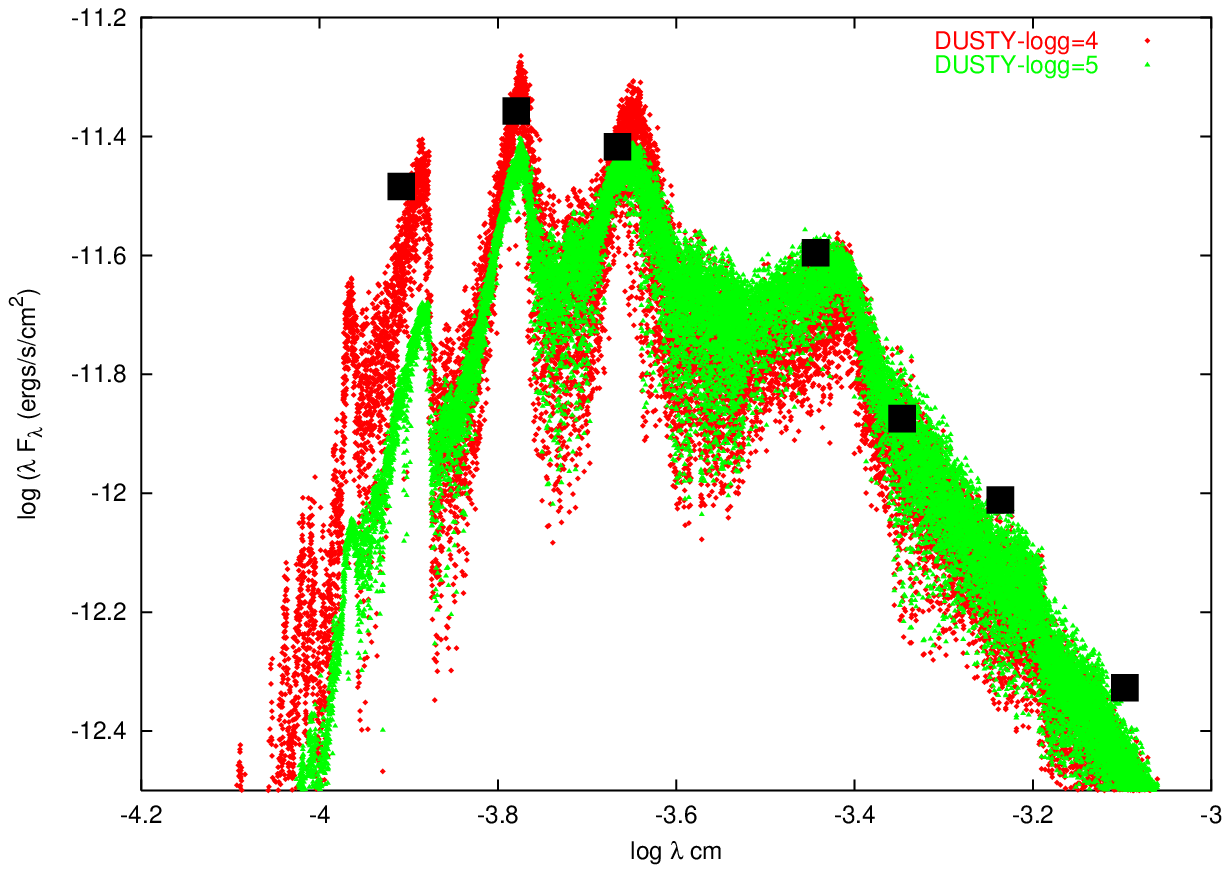}} \\
      \resizebox{80mm}{!}{\includegraphics[angle=270]{f1c.epsi}} \\
    \end{tabular}
    \caption{{\it Top} (a) DUSTY (blue), SETTLE (green) and COND (red) model fits for 2M1315. JHK magnitudes are from 2MASS, IRAC from this work. The errors in IRAC magnitudes are between 0.02 and 0.06 mag, and are smaller than the symbol size.
    {\it Middle} (b) DUSTY model fits for log g=4 (red) and log g=5 (green) for a $T_{eff}$ of 1700 K.
    {\it Bottom} (c) Near-IR CMD comparing the three models: {\it solid and long-dashed lines}, DUSTY and COND models for 1 Gyr, respectively; {\it short-dashed line}, SETTLE models mapped onto the DUSTY 1 Gyr $T_{eff}$-radius relation (from Allard et al. 2003). {\it Filled square}, 2M1315; {\it open squares, filled triangles and crosses}, M, L and T dwarfs from Dahn et al. (2002), respectively.}
  \end{center}
\end{figure}

\clearpage
\begin{figure}
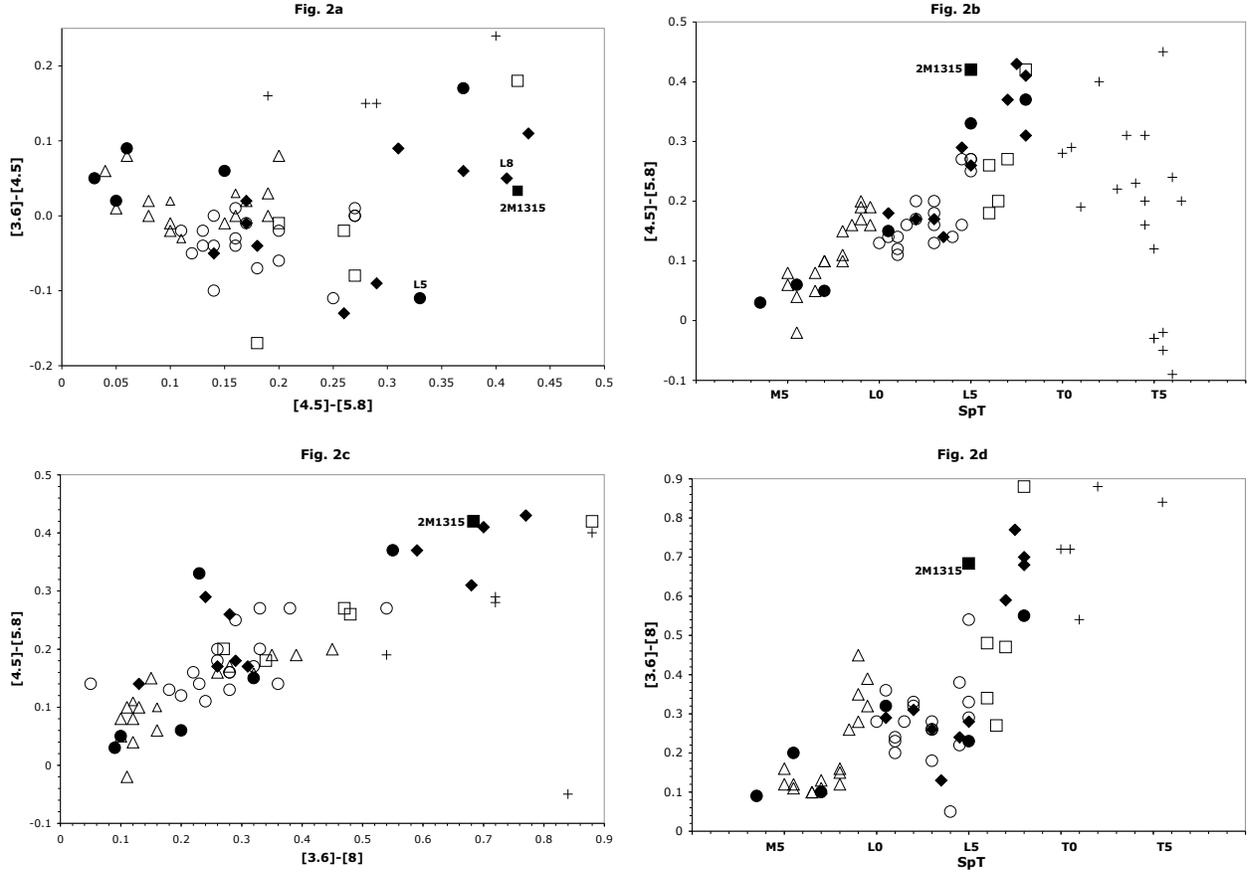

 \begin{center}
    \begin{tabular}{cc}
      \resizebox{80mm}{!}{\includegraphics[angle=270]{f2a.epsi}} &
      \resizebox{80mm}{!}{\includegraphics[angle=270]{f2b.epsi}} \\
      \resizebox{80mm}{!}{\includegraphics[angle=270]{f2c.epsi}} &
      \resizebox{80mm}{!}{\includegraphics[angle=270]{f2d.epsi}} \\
    \end{tabular}
    \caption{IRAC colors for 2M1315, along with M, L and T dwarfs from Patten et al. (2006). {\it Filled square}, 2M1315; {\it open triangles}, M dwarfs; {\it open circles}, L0-L5; {\it open squares}, L6-L8; {\it plus signs}, T dwarfs; {\it filled diamonds}, L dwarfs with Li EW measurements available; {\it filled circles}, M and L dwarfs with $V_{tan}$ between 50 and 80 km/s. }
  \end{center}
\end{figure}

\clearpage
\begin{figure}
\resizebox{150mm}{!}{\includegraphics[angle=270]{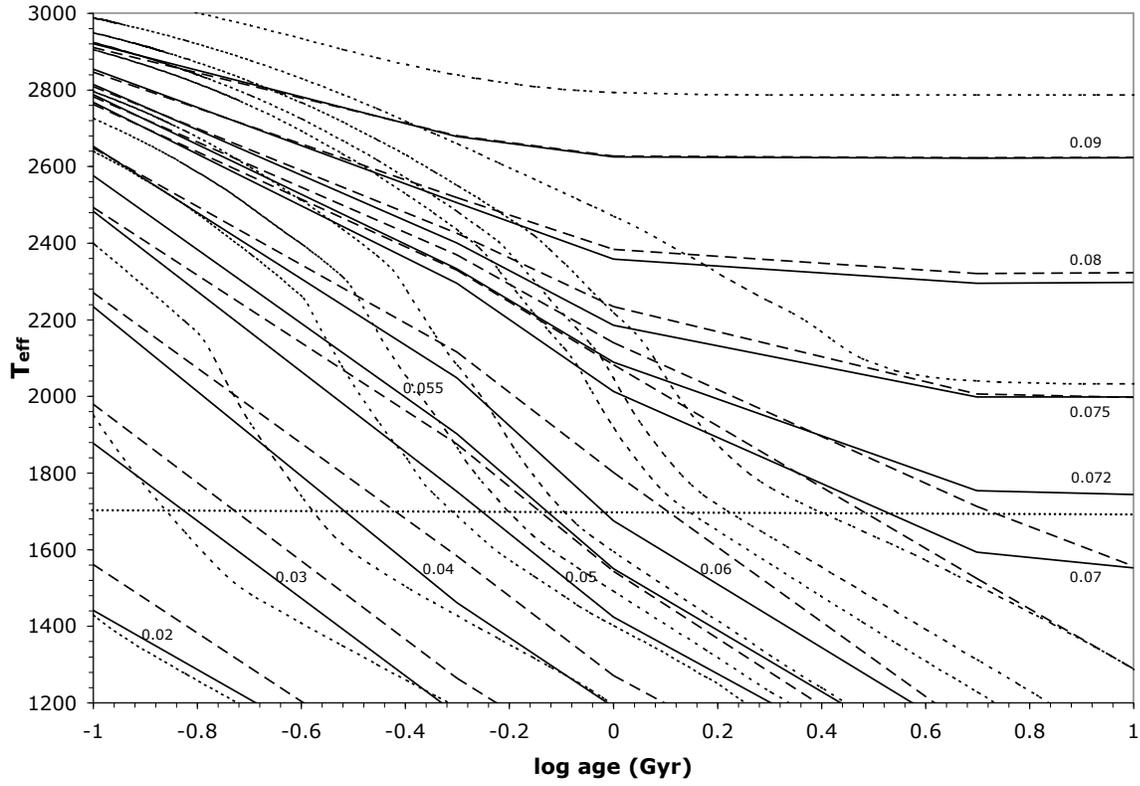}}
\caption{Evolutionary tracks for masses 0.09$M_{\sun}$ (top) to 0.02$M_{\sun}$ (bottom). {\it Solid lines}, DUSTY (C00); {\it long-dashed lines}, COND (C00); {\it short-dashed lines}, B97. }
\end{figure}

\clearpage
\begin{deluxetable}{ccccccccc}
\tabletypesize{\scriptsize}
\tablecaption{IRAC Observations}
\tablewidth{0pt}
\tablehead{
\colhead{Star} & \colhead{3.6 $\micron$} & \colhead{S/N} & \colhead{4.5 $\micron$} & \colhead{S/N} &
\colhead{5.8 $\micron$} & \colhead{S/N} & \colhead{8 $\micron$} &
\colhead{S/N} \\
\colhead{} & \colhead{mJy} & \colhead{} & \colhead{mJy} & \colhead{} & \colhead{mJy} &
\colhead{} & \colhead{mJy} & \colhead{}  
}
\startdata
2M1315	&	3.10$\pm$0.06 	&	60.86	&	2.05$\pm$0.04	&	60.19	&	1.92$\pm$0.02	&	119.53	&	1.33$\pm$0.02	&	57.32 \\
\enddata
\tablecomments{The 1-$\sigma$ errors do not include IRAC calibration uncertainty of $\sim$2\%.}
\end{deluxetable}

\begin{deluxetable}{cccc}
\tabletypesize{\scriptsize}
\tablecaption{Six epochs of spectroscopy for 2M1315}
\tablewidth{0pt}
\tablehead{
\colhead{Date of Observation}  & \colhead{H$\alpha$ EW ($\AA$)} & \colhead{Spectral Resolution($\AA$)} & \colhead{Ref} 
}
\startdata
2001 March 30  & 121 $\pm$ 31 & 8.6 & Hall (2002a) \\
2001 May 1-2  & 97$\pm$10 & 2.8 & Gizis (2002) \\
2001 August 15  & 25 $\pm$ 10 & 8.6 & Hall (2002a) \\
2002 March 15  & 24.1 & 0.1 & Fuhrmeister et al. (2005)  \\
2002 September 9  & 124 $\pm$ 55 & 4.6 & Hall (2002b) \\
2003 March 11 & 153 $\pm$ 26 & 2.7 & Barrado y Navascu\'{e}s (2006) \\
\enddata
\end{deluxetable}


\begin{thebibliography}{}
\bibitem[Allard et al. 2001]{all01} Allard, France, Hauschildt, Peter H., Alexander, David R., Tamanai, Akemi \& Schweitzer, Andreas, 2001, \apj, 556,357
\bibitem[Barrado y Navascu\'{e}s 2006]{b06} Barrado y Navascues, D., 2006, astro-ph/0608478
\bibitem[Berger 2006]{ber06} Berger, E., 2006, astro-ph/0603176
\bibitem[Burgasser et al. 2000]{bur00} Burgasser, Adam J.; Kirkpatrick, J. Davy; Reid, I. Neill; Liebert, James; Gizis, John E.; Brown, Michael E., 2000, \aj, 120, 473
\bibitem[Burrows et al. 1997]{b97} Burrows, A. et al. 1997, \apj, 491, 856 (B97)
\bibitem[Chabrier et al. (2000)]{ch00} Chabrier, G., Baraffe, I., Allard, F., \& Hauschildt, P., 2000, \apj,542, 464 (C00)
\bibitem[Cushing et al. (2006)]{cu06} Cushing, Michael C. et al. 2006, \apj, 648, 614
\bibitem[Dahn et al. 2002]{dahn02} Dahn, Conard C.; Harris, Hugh C.; Vrba, Frederick J.; Guetter, Harry H.; Canzian, Blaise; Henden, Arne A.; Levine, Stephen E.; Luginbuhl, Christian B.; Monet, Alice K. B.; Monet, David G.; and 8 coauthors, 2002, \aj, 124, 1170
\bibitem[Fazio et al.(2004)]{irac} Fazio, G., et al.  2004, \apjs, 154, 10
\bibitem[Fuhrmeister et al. 2005]{fuhr05} Fuhrmeister, B., Schmitt, J. H. M. M., Hauschildt, P. H., 2005, \aap, 439, 1137
\bibitem[Gizis 2002]{gizis02} Gizis, J. E., 2002, \apj, 575, 484
\bibitem[Gizis et al. 2000]{gizis00} Gizis, John E.; Monet, David G.; Reid, I. Neill; Kirkpatrick, J. Davy; Liebert, James; Williams, Rik J., 2000, \aj, 120, 1085
\bibitem[Hall 2002a]{hall02a} Hall, P. B., 2002a, \apjl, 564, L89
\bibitem[Hall 2002b]{hall02b} Hall, P. B., 2002b, \apjl, 580, L77
\bibitem[Hauschildt et. al.(1999a)]{hau99a} Hauschildt, P. H., Allard, F., \& Baron, E. 1999a, \apj, 512, 377 
\bibitem[Hauschildt et. al.(1999b)]{hau99b} Hauschildt, P. H., Allard, F., Ferguson, J., Baron, E., \& Alexander, D. R. 1999b, \apj, 525, 871
\bibitem[Kirkpatrick 2005]{kir05} Kirkpatrick, J. D., 2005, A\&AAR, 43, 195
\bibitem[Lada et al. (2006)]{lada06} Lada, C. J. et al. 2006, \aap, 131, 1574
\bibitem[Liebert et al. (1999)]{lie99} Liebert, J. et al. 1999, \apj, 519, 345
\bibitem[Liebert et al. 2003]{lie03} Liebert, James; Kirkpatrick, J. Davy; Cruz, K. L.; Reid, I. Neill; Burgasser, Adam; Tinney, C. G.; Gizis, John E., 2003, \aj, 125, 343
\bibitem[Marley et al. 2002]{m02} Marley, M. S. et al. 2002, \apj, 568, 335
\bibitem[Mohanty et al. (2002)]{moh02} Mohanty, S. et al. 2002, \apj, 571, 469
\bibitem[Mullan \& MacDonald 2001]{mm01} Mullan, D. J., \& MacDonald, J., 2001, \apj, 559, 353
\bibitem[Noll et al. 2000]{noll00} Noll, Keith S., Geballe, T. R., Leggett, S. K., \& Marley, Mark S., 2000, \apjl, 541, L75
\bibitem[Oppenheimer et al. 1998]{opp98} Oppenheimer, B. R.; Kulkarni, S. R.; Matthews, K.; van Kerkwijk, M. H., 1998, \apj, 502, 932
\bibitem[Patten et al. (2006)]{pa06} Patten, B. M. et al., astro-ph/0606432
\bibitem[Reid et al. (1999)]{re99} Reid, N. et al. 1999, \apj, 527, L105
\bibitem[Reid 2003]{reid03} Reid, N., 2003, \mnras, 342, 837
\bibitem[Reiners \& Basri (2006)]{rb06} Reiners, A., \& Basri, G., 2006, astro-ph/0610365
\bibitem[Roellig et al. 2004]{roe04} Roellig, T. L.; Van Cleve, J. E.; Sloan, G. C.; Wilson, J. C.; Saumon, D.; Leggett, S. K.; Marley, M. S.; Cushing, M. C.; Kirkpatrick, J. D.; Mainzer, A. K.; Houck, J. R., 2004, \apjs, 154, 418
\bibitem[Schneider et al. (1991)]{sch91} Schneider, D. P. et al. 1991, \aj, 102, 1180
\bibitem[Tsuji et al. 1996]{tsu96} Tsuji, T., Ohnaka, K., Aoki, W. \& Nakajima, T., 1996, \aap, 308, L29
\bibitem[Werner et al.(2004)]{spitzer} Werner, M., et al.  2004, \apjs, 154, 1
\bibitem[White \& Basri (2003)]{wb03} White, R. J. \& Basri, G. 2003, \apj, 582, 1109

\end{thebibliography}
\end{document}